# SYM: Toward a New Tool in User's Mood Determination

**Willy Yvart**
DeVisu (UVHC)
TCTS (UMONS)
Rue Michel Rondet,
59135 Wallers, FRANCE
willy.yvart@neuf.fr

**Charles-Alexandre Delestage**
DeVisu (UVHC)
Rue Michel Rondet,
59135 Wallers, FRANCE
ca.delestage@gmx.fr

**Sylvie Leleu-Merviel**
DeVisu (UVHC)
Rue Michel Rondet,
59135 Wallers, FRANCE
sylvie.merviel@univ-valenciennes.fr

## ABSTRACT
Even though the emotional state is increasingly taken into account in scientific studies aimed at determining user experience of user acceptance, there are still only a few normalized tools. In this article, we decided to focus on mood determination as we consider this affective state to be more pervasive and more understandable by the person who is experiencing it. Thus, we propose a prototypical tool called SYM (Spot Your Mood) as a new tool in user mood determination to be used in many different situations.

## ACM Classification Keywords
Algorithms; Experimentation; Human Factors; Measurement; Theory

## INTRODUCTION
Since Aristotle and the Greek philosophers, the question of the affective state has stayed central and very questioned in most areas of science. With the emergence of psychology and other cognitive sciences, the subject seems to be even more unavoidable. The constructivist paradigm, by putting the person in the very center of the sense-making process, collaterally brings the question of the influence of his affective state to the table [1].

Even though the subject has been treated over hundreds of years, it is quite noticeable that it still suffers from a lack of consensus and sometimes a lack of recognition in most technical areas. Actually, there is a lot of confusion; most terms and concepts are ill-defined and they often cannot escape from the epistemic posture of authors.

In this article, we decided to work on the mood concept that is indisputably less "fashionable" that the emotion concept. We put forward that mood is, however, a central parameter of the whole configuration of mental activity. Having said that, the question of knowing about the mood of the person or the population we are studying in an experimentation cannot remain unanswered, it has to be studied as a potential parameter of the experimental situation in order not to become a bias. Thus, after disambiguating as best as possible the concept of mood, we propose a protocol and a prototypical tool intended to solve that question.

## DETERMINATION OF AN AFFECTIVE STATE

### Why mood?
When taking into account the emotional state (as opposed to traits we are not talking about here) in determining its effect on reasoning or decision making, we first need to disambiguate which affective concept fits the most with user experience considerations.

Empirically, when confronted with some experience, listening to music, visiting a museum and so on, a person induces a change in his actual affective state. Nonetheless, the qualification of this affective state is still quite fuzzy. Generally, in the literature, we can find three main concepts that are sometimes mixed up: "affect", "mood" and "emotion".

First, "affect" seems to be a portmanteau word, a concept embodying "mood", "emotion", "feelings", etc. There is, at the same time, confusion with the word "affect" in itself. As used in the expression "affective state", it refers to a state, an immediate emotionally "colored" cognitive state, but, the word can also be used to qualify feelings in the most sensitive signification. Then, as "affect" appears to be a supra-concept covering different ill-defined infra-concepts, we have to rely on another.

The concept of emotion, even if it is the most commonly encountered in the literature, also seems to be the fuzziest. In fact, more than 30 years ago, Kleinginna and Kleinginna were able to point out more than one hundred different definitions [2]. By researching for this communication we noticed over two hundred definitions that are not compatible: 38% of these consider emotions as being a physiological phenomenon while 62% consider them as being a psychological one. This lack of consensus brings about most of the misunderstandings we can notice in the literature. Nevertheless, there is some agreement; in most definitions "emotion" refers to a response to some internal or external stimulation of short-term duration and quite strong effects on the behavior. Thus, emotions are seen as being intentional (associated to an object or a stimulus) and transient. By looking again in the literature (the articles in which we found the definitions, which numbered over 200) we could see that a third of the statements we found considered that emotions are inaccessible to the conscious mind, versus two thirds which advocated the opposite. To us, emotion seems to be, most of the time, a "*qualia-like*" phenomenon being analyzed as a part of the information available in order to copy with the situation. The flow of data of a physiological nature (heart rate, blood pressure, etc.) or psychological nature (feelings) is melted into the rest of the sensitive and cognitive information during the experience. That implies that emotions are too volatile, too transient and not stable on a macroscopic time scale. According to Sloboda [3], emotions are "rather unmemorable on average";even though this statement looks polemical, it is still possible to agree to with it despite some reservation. Strong reciprocal links between emotions and memory have been shown from the recording to the reading of the information, and these cannot be denied [24-26]. However, and in accordance with Sloboda's statement it is not the emotional experience in itself that is recorded or recalled. We can remember about contextual cues, ongoing actions (like in the Ziegarnik's effect [27]) and about the consequences of the experience. We do record/recall the initial setup (initial mood, environment, etc.), the perturbing element, and the final setup (consequences, final mood,..) with some kind of an emotional label on it which is like the very source of feelings such as nostalgia for example [28]. Body feelings, sensations, *etc.* are, to a large extent, lost as it is for any "*qualia-like*" phenomenon.

So, in an attempt, like ours, to try to characterize the affective experience of a person, they constitute a cul-de-sac.

As a consequence, our epistemology has to be built on the last commonly encountered concept: the "mood". The concept is sometimes considered as being old fashioned, especially in its French translation "*humeur*". Even if we could notice up to 88 different statements in the literature, chronologically from Weld [4] to Juslin and Sloboda [5], there is more agreement between authors.

Hence, the mood stands for a state, at a given time, in a given situation, that is linked to the framework of the whole organism by the attention given to its sense-making and cognitive processes [6], [7]. Thus, mood is seen as the psycho-physiological substratum of any situation; whenever we are in a certain situation - in other words, all the time - we are in a certain mood [8] (very similar to Heidegger's concept of "*stimmung*" [9]) even though we are only able to be conscious of that state if we focus our attention on it.

Mood can be seen as a relatively stable and pervasive setup that can be maintained from minutes to days. Without a perturbing element, a mood can be sustained or can slowly decay to a more "stable" one. We can draw a quick analogy with a homeostatic state in the chemical sciences. In the attempt of Lane and Terry to give a theoretical basis to the concept of mood, they put forward that it is "*a set of feelings, ephemeral in nature, varying in intensity and duration, and usually involving more than one emotion*" [8]. In this context, emotions can then be seen as the perturbing elements we were talking about. This posture is compatible with the fact that emotions are quasi-inaccessible in themselves to the cognitive process. In our opinion, it is the mood and the difference between a pre-stimulation and a post-stimulation mood that are accessible to the cognitive process.

**How to measure/represent mood?**
That being said, there is still a central question to answer: how to measure mood and how to represent it in a sufficiently efficient and communicable way? Following on from what we just said, we can put as a working hypothesis that we are trying to characterize an affective experience in a certain situation by looking at the mood and at its evolution during this situation. Although emotions are not accessible, we still can notice the "dents" they produce on the mood.

So, we need to follow the evolution of an internal parameter of the visitor / listener / spectator. Nevertheless, to follow an evolution, we need to characterize differences. In fact, we need to find information about the quasi-homeostatic affective state at the beginning of the experience we want to study and, *a minima*, the final one. Thus, the difference between the two "measures" would be analogous to a kind of integrative view of the affective experience.

But, that does not answer the "how to" question. At that time, we have to choose between two epistemic points of view which are opposite to each other. On one hand, many psychologists and scientists put forward that it is possible to measure some affective phenomenon by looking at physiological reactions like micro-expressions [10] or heart-rate[11]. On the other hand, we can emphasize Barrett's statement saying: "*for better or worse, self-report represents the most reliable and possibly only window that researchers have on conscious, subjective emotional experience.*" [12]

Taking a look at Mugur-Schächter's works on theorizing the qualification and measuring process [13], [14], we would preferably agree with Barrett. Actually, to qualify implies to measure and to measure implies a measuring tool. Whatever we are trying to qualify, we first need to build a link to it starting from the conscious process. Something from the outside of our body will then be measured throughout our biosensors, our attention and our cognition. Something from the inside of our body will be measured throughout our attention to it and our cognition.

But it seems common grounded that affective phenomena like mood are both internal and external phenomena from a conscious point of view. Consequently, looking for somebody else's mood by looking at physical or physiological variation is like erasing a big part of the information. The only person able to collect information from the situation and from its inner world is then the person living the experience.

Nevertheless, that statement does not totally disqualify physiological measures like heart-rate. Actually, heart-rate variability can be a useful additive signal that can constitute a highly profitable source of information.

**Asking for the mood**

If the person is in the best position to determine his mood, we now have to ask him for it. According to Shannon's theory of information, a transmitter and a receiver have to speak (at least) the same language to code and decode the message. That is why spoken or written language appeared to be a good idea at their origin.

But, as we were able to notice during previous experimentation, verbalization is not the easiest way to study affective response when listening to music [15]. Actually, in an experiment we asked for people to qualify the mood of a musical excerpt using adjectives. We wanted to test an open-ended condition in order to estimate personal judgment about the music expressed in their own language.

The supra-context of the experiment was to try to establish a minimal set list of mood adjectives in order to use them as the standing ground for indexing and searching for the musical excerpt in an online library (the reason why we needed to have the mood expressed in the user's own language). In this context, the qualifier has to reach a certain threshold of consensus in order to be meaningful. Due to the impressive quantity of different adjectives we gathered, we then decided to build up some semantic proxemy analysis so as to meet an inter-subjective agreement.

In this attempt, we discovered that there was agreement to a certain degree. Actually, most of the terms seemed to coincide with a main idea as they were, for the most part, synonyms. But we could also notice that there was some confusion. Indeed, there appeared to be a discrepancy between the personal mood of the person and an extra-personal mood designating the excerpt. That sounds logical to say that in the musical signal there is no mood, only a living person can endure such an affective state, but when submitted to a *stimulus* able to induce or to produce emotions, a person is able to assign an emotional content to that *stimulus*. At that moment, there can be confusion between a "personally-felt" mood and a "perceived" one [16], [17].

The results of this experiment brought us to develop a new protocol to deal with the issues of verbalization and confusions between moods.

**SYM: SPOT YOUR MOOD**

Getting over the verbalization problem is a big issue. In the context of our first experiment, deleting the massive set of qualifiers would represent a more efficient tagging process resulting in a better indexation. Thus, we decided to rely on some graphical expression and representation systems to assess for the mood. Like in the Self-Assessment Manikin [18] or in the Geneva Emotion Wheel [19] everything starts with the use of dimensional theory of emotion and the valence-arousal space notably developed by Russell [20] and Thayer [21].The valence dimension consists in a hedonic value of pleasantness / unpleasantness and the arousal dimension is relative to a psycho-physiological wakefulness. According to Russell, every possible affect is representable by a set of two values in those dimensions.

The SAM allows the user to give information about his emotional state through valence, arousal and control/potency (the last is not always represented) discrete 10-step-Likert-scales. The GEW consists in a flower of circles representing the emotional universe cut into discrete families of emotions. Both give a discrete representation of the mood space. In our approach, we decided to stay more faithful to the dimensional approach described by Russell.

Thus, we directly present a valence-arousal space to the user and ask him to express his mood throughout one or many points. In our representation, the valence dimension is the abscissa axis and the arousal is the ordinate axis. *Extrema* are then characterized with smileys representing high valence, low valence, high arousal and low arousal in order to avoid verbal influence.

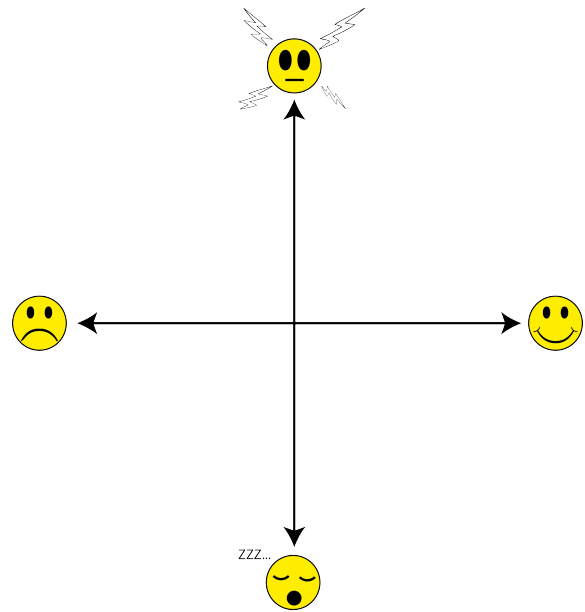

**Figure 1 : SYM's Valence-Arousal diagram**

Until then, the user is free at any time to designate a point on the figure to give us information on mood. We let the user have the impression of a continuous scale, however the indications are digitally stored in the discretized interval[- 100;100][-100;100]. The software architecture is a traditional client-server routine, but the idea is to allow the researcher to use plug-ins such as indoor localization or synchronization with audio or video, depending on the experimental needs.

Also, the major utility of SYM is to provide a dynamic dictionary of emotional states linked to valence-arousal locus. This dictionary consists in a database, which links nouns, verbs, adverbs and expressions related to the same emotional concept. With semantic proxemy analysis between the terms, it is possible to also link some emotional states in order to create a kind of net.

The user spots a mood which is represented as a couple of (x,y) coordinates. These coordinates are transmitted to a server, which interrogates a data-base and finds the 3 closest mood terms and sends them to the client. If the user is satisfied, the server stores the tuple (x, y, word). If not, the process is the same, excepting the last words refused by the client. Doing that, SYM can be seen as a verbalization helper. As we were able to notice during previous experimentations, it is quite hard to express clearly how we feel when asked to do so. Moreover, direct verbalization also presents polysemic issues and the use of adjectives that a computer cannot disambiguate.

Here, as every situation can lead to the usage of the same words with different meanings, a section of the master dictionary can be set up to fit the expression of the users. Also, some words will not be used on some occasions. For example, you will rarely find "anger" expressed as a musical mood. Thus, with the help of some specialists, for example in museography, we can determine the terms that would fit or not to the situation, then generating custom dictionaries. These custom dictionaries bring more simplicity for the user, avoiding "noise words" being proposed while spotting on the diagram.

Periodically, the server makes a separate update of the position of words on the valence-arousal based on the user's feedback, to include the folksonomy generated throughout the different experimentations. As a result, the dictionary used for each setup is a part of the master dictionary with custom placements of the mood words regarding the conditions.

This function allows us to make a difference for the usage of words, which can differ for certain activities (shopping, cultural exhibitions,…) or within different populations (street language, rich language, etc.).

The architecture permits us to use the client terminal or to visualize in a browser the results of the experiments, placed on the diagram for each user or in a cloud of points. The raw data can also be imported as *.csv* files on any statistics software for treatment and synchronization with physiological signals or experiment props (music, video…)

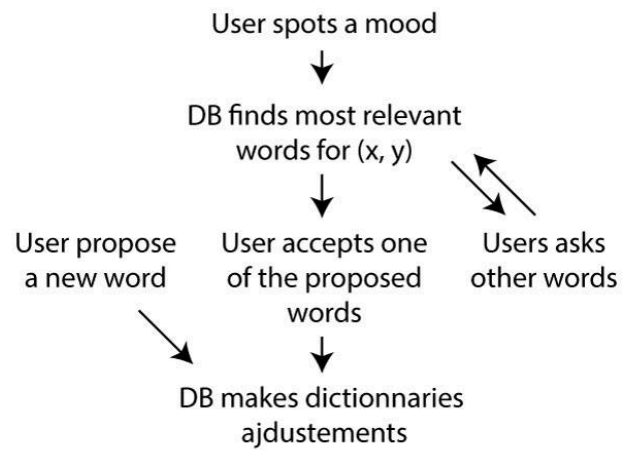

**Figure 2 : SYM simplified workflow**

## MOOD SPOTTING

### A validation of mood spotting on Valence-Arousal space

The usage of a valence-arousal space with a similar approach can be found on the experiments led by Fanny Bougenies during her PhD thesis [23]. The protocol was using it to assess the emotional state of about 130 children with or without deficiencies (deaf, mental impairment, autism…) before and after a visit to a museum. Each of them was given a tablet with an educational software program designed to enhance their visit. At that time, valence-arousal graphs were paper-made and consisted in a part of the questionnaire.

The conclusions showed that not a single child endured difficulties understanding, they managed to use the diagram without indications. Moreover, it allowed autistic children, generally not willing to talk much, to express their emotions smoothly.

The problem was that all these indications were taken on a calibrated sheet of paper. It meant precise measurement, normalization and reporting on a computer (with a computer-aid design software) to get a usable corpus of data. This underlined the necessity of finding an appropriate computerized workflow to gather and process the moods.

### Mood spotting for determining visitor experience in museum exhibition

The testing experimentation was focused on the ability of the interface to be used by the public without altering their experience with a significant impact. It took place in the "Musée des Beaux-Arts" of Lille in France. The visitors wanting to visit the museum were offered a free pass just before the act of payment for their tickets, but they were required to be equipped with eye-tracking glasses and a tablet with an android version of SYM. At any time, they could spot their mood on the device. For 50% of them, a word was also provided to qualify their mood, based on the point they spotted, which they had to validate or not. After their visit, they were shown their field of view during their visit (from the eye-tracking glasses) and were asked to describe their experience.

The result was that all of the interviewees were able to spot their mood easily, and without any long explanations of the system. Also, the impressions and comments given after the visit correlated with the tuples (x, y, word) spotted on the application. If not as precise as a full interview after the experiment, the SYM approach was able to give the shape of the experience of the visitor just as the affective state shapes the user experience. This could be the first step of an evaluation protocol to distinguish the points in the exhibition that have issues.

**CONCLUSION**

SYM was originally designed as a "verbalization" helping tool when it comes to expressing mood. Even if we are still prototyping the tool in itself, we managed to build a complete integrative workflow that we are still amending. The experiment we presented showed us that SYM was very efficient as a non-pervasive experimental protocol when inquiring about affective states.

The participants were only told to spot their mood at the beginning and at the end of the visit and then were free, for the rest of time, to inform us or not about their state of mind. This results in a very small probability of being exposed to the social desirability bias [22] which could be feared as we penetrate into the very private area of affective states.

Before the experiment took place, we had already noticed the efficiency of pre-SYM attempts. In a still unpublished study about qualifying music throughout mood spotting, we noticed that the listeners were able to understand the diagram at a glance and give a point for their mood.

We also noticed that users were able not to mistake between their own mood and the musical one. Actually, in order not to fall into this trap, we decided to inquire into their mood before and after the listening and about the "musical" mood during the listening. It is interesting to see that there is no possible inter-personal common-ground about the own-mood (nor about its evolution) although we can recognize a certain degree of convergence for the inter-personal musical moods. That can be seen as a successful distinction between the personal and the musical mood. Furthermore, that gave us two interesting pieces of information about the music; on one hand, the difference between the beginning and the end of the experiment related to the emotional effect of the music and, on the other, a mood assigning for the "emotional content" of the musical excerpt in itself.

Admittedly, SYM is still a prototype and it needs to be improved and amended to fit the different experimental setups, nevertheless, according to the first results we had in the field, SYM can open new perspectives in user mood determination.

**ACKNOWLEDGMENTS**

SYM is a part of a co-directed program between the university of Mons in Belgium and the university of Valenciennes in France. We would like to thank Caroline Guillaume-Blaydes for her corrections to our text and all the participants in the different experiments we are still working on.

**REFERENCES**
1. H. A. Simon, 'Motivational and Emotional Controls of Cognition', *Psychol. Rev.*, vol. 74, no. 1, pp. 29–39, 1967.
2. P. R. J. Kleinginna and A. M. Kleinginna, 'A categorized list of emotion definitions, with suggestions for a consensual definition', *Motiv.Emot.*, vol. 5, no. 4, pp. 345–379, 1981.
3. J. A. Sloboda, 'Music in everyday life : the role of emotions', in *Handbook of music and emotion : theory, research, applications*, Oxford, England: Oxford University Press., 2011, pp. 493–514.
4. H. P. Weld, 'An experimental study of musical enjoyment', *Am. J. Psychol.*, vol. 23, no. 2, pp. 245–308, 1912.
5. P. N. Juslin and J. A. Sloboda, *Handbook of music and emotion : theory, research, applications*. Oxford: Oxford University Press, 2011.
6. J. P. Forgas, 'Mood and Judgment: The Affect Infusion Model (AIM)', *Psychol. Bull.*, vol. 117, no. 1, pp. 39–66, 1995.
7. J. D. Green, C. Sedikides, J. A. Saltzberg, J. V. Wood, and L.-A. B. Forzano, 'Happy mood decreases self-focused attention', *Br. J. Soc. Psychol.*, vol. 43, pp. 147–157, 2003.
8. A. M. Lane and P. C. Terry, 'The nature of mood : development of a theoretical model', *J. Appl. Sport Psychol.*, vol. 12, no. 1, pp. 16–33, 2000.
9. M. Heidegger, *Being and Time: A Translation of Sein und Zeit*. SUNY Press, 1996.
10. P. Ekman, 'Facial expression and emotion', *Am. Psychol.*, vol. 48, no. 4, pp. 376–379, 1993.
11. F. Bousefsaf, C. Maaoui, and A. Pruski, 'Remote assessment of the heart rate variability to detect mental stress', in *Proceedings of the 7th International Conference on Pervasive Computing Technologies for Healthcare*, 2013, pp. 348–351.
12. L. F. Barrett, 'Hedonic tone, perceived arousal, and item desirability : Three components of self-reported mood', *Cogn. Emot.*, vol. 10, no. 1, pp. 47–68, 1996.
13. M. Mugur-Schächter, *Sur le tissage des connaissances*. Paris, France: Hermès, 2006.
14. M. Mugur-Schächter, 'Les leçons de la mécanique quantique', *Le débat*, vol. 2, no. 94, pp. 169–192, 1997.
15. W. Yvart, T. Dutoit, and S. Dupont, 'Une approche info-communicationnelle des librairies musicales en ligne', in *Proc. SFSIC'14*, Toulon, France, 2014.

16. L.-O. Lundqvist, F. Carlsson, P. Hilmersson, and P. N. Juslin, 'Emotional Responses to Music : Experience, Expression, and Physiology', *Psychol. Music*, vol. 37, no. 1, pp. 61–90, 2009.
17. P. N. Johnson-Laird and K. Oatley, 'Emotions, Music, and Litterature', in *Handbook of emotions*, 3rd Edition., New York, NY, USA: Guilford Press, 2008, pp. 102–113.
18. M. M. Bradley and P. J. Lang, 'Measuring emotion: the self-assessment manikin and the semantic differential', *J. Behav. Ther. Exp. Psychiatry*, vol. 25, no. 1, pp. 49–59, 1994.
19. K. R. Scherer, 'What are emotions? And how can they be measured?',*Soc. Sci. Inf.*, vol. 44, no. 4, pp. 695–729, 2005.
20. J. A. Russell, 'Affective Space is bipolar', *J. Pers. Soc. Psychol.*, vol. 37, no. 3, pp. 345–356, 1979.
21. R. E. Thayer, *The Biopsychology of Mood and Arousal*. New York, NY, USA: Oxford University Press, 1989.
22. D. P. Crowne and D. Marlowe, 'A new scale of social desirability independent of psychopathology.',*J. Consult. Psychol.*, vol. 24, no. 4, p. 349, 1960.
23. F. Bougenies, 'Expérience de visite muséale for all : visite augmentée et construction de sens. Le cas d'enfants avec et sans handicaps au Palais des Beaux-Arts de Lille.', *Université de Valenciennes et du Hainaut-Cambrésis*, 2015
24. G. H. Bower, S. G. Gilligan, and K. P. Monteiro, 'Selectivity of learning caused by affective states', *J. Exp. Psychol. Gen.*, vol. 110, no. 4, pp. 451–473, 1981.
25. J. R. Anderson and G. H. Bower, *Human associative memory*. Washington, D.C., USA: Winston & Sons, 1973.
26. S. Jhean-Larose, N. Leveau, and G. Denhière, 'Influence of emotional valence and arousal on the spread of activation in memory', *Cogn. Process.*, pp. 1–8, 2014.
27. B. Ziegarnik, 'Das Behaltenerledigter und unerledigter Handlungen', *Psychol. Forsch.*, vol. 9, pp. 1–85, 1297.
28. M. M. Bradley, M. K. Greenwald, M. C. Petry, and P. J. Lang, 'Remembering pictures: pleasure and arousal in memory.',*J. Exp. Psychol. Learn. Mem. Cogn.*, vol. 18, no. 2, p. 379, 1992.